\documentclass[twocolumn]{revtex4}
\usepackage{amsmath,dcolumn,feynmf,epsfig,CJK}
\begin{document}
\title{ON THE QUANTIZED RELATIVISTIC MEAN FIELD THEORY\\ FOR NUCLEAR MATTER
\footnote{Supported by the Nature Science Foundation of China (Grant
Nos. 10875003 \& 10811240152). And the calculations are supported by
CERNET High Performance Computing Center in China.} }
\author{QI-REN ZHANG 
}
\affiliation{School of Physics, Peking University, Beijing 100871,
People's Republic of China}
\author{CHUN-YUAN GAO 
\footnote{E-mail: gaocy@pku.edu.cn}} \affiliation{School of Physics
and State Key Laboratory of Nuclear Physics and Technology, Peking
University, Beijing 100871, People's Republic of China}
\begin{abstract}
We propose a quantization procedure for the nucleon-scalar meson
system, in which an arbitrary mean scalar meson field $\phi$ is
introduced. The equivalence of this procedure with the usual ones is
proven for any given value of $\phi$. By use of this procedure, the
scalar meson field in the Walecka's RMFT and in Chin's RHA are
quantized around the mean field. Its corrections on these theories
are considered by perturbation up to the second order. The
arbitrariness of $\phi$ makes us free to fix it at any stage in the
calculation. When we fix it in the way of Walecka's RMFT, the
quantum corrections are big, and the result does not converge. When
we fix it in the way of Chin's RHA, the quantum correction is
negligibly small, and the convergence is excellent. It shows that
RHA covers the leading part of quantum field theory for nuclear
systems and is an excellent zeroth order approximation for further
quantum corrections, while the Walecka's RMFT does not. We suggest
to fix the parameter $\phi$ at the end of the whole calculation by
minimizing the total energy per-nucleon for the nuclear matter or
the total energy for the finite nucleus, to make the quantized
relativistic mean field theory (QRMFT) a variational method.


{\bf PACS number(s)}:03.65.Ca, 21.60.-n, 21.65.-f
\end{abstract} \maketitle
\section{Introduction}
Relativistic Mean Field Theory (RMFT) is a fruitful and widely used
theory in nuclear
physics\cite{l,w74,c2,sw,w,fps,c1,bs83,z2,r1,r2,b,z3,g}. Its
agreement with observational data is impressive. It seems to show
that the nuclear data under certain energy scale, say several
hundred MeV, may be roughly understood by hadron field theory. Of
course, it should be a quantum theory of fields. But in RMFT, meson
field operators are replaced by their expectation values, which are
considered as classical. RMFT is therefore semi-classical. A
meaningful RMFT must be followed by quantum corrections due to meson
field quantization, and the correction should not qualitatively
change its agreement with observational data. In this case, one may
realize a quantum hadron field theory for nuclear systems under
certain energy scale, with RMFT to be the zeroth order contribution,
and calculate quantum corrections by perturbation.

Quantum hadron field theory has already been developed for nuclear
physics in the usual loop expansion
formalism\cite{l,w74,c2,sw,w,fps}, in which the mean meson field is
the contribution of the tadpole diagrams. The attachment of nucleon
loops (tadpole heads) on a nucleon line by the scalar meson lines
(tadpole tails) changes the nucleon mass. Since the tadpole head
itself is also formed by a nucleon line, the RMFT calculation for a
nuclear system means an infinite series of attachments of tadpoles
on tadpoles. In this sense, RMFT is non-perturbative, like the
method of self-consistent field widely used in quantum mechanics. A
reasonable way of considering its quantum corrections is to quantize
the field around its classical solution, like the consideration of
residual interactions in many-body problems on the basis of its
self-consistent field solutions, or like the consideration of
quantum corrections in laser-atom interaction on the basis of its
semi-classical solutions\cite{z}. This is a quantization procedure
for nucleon field expanded on a complete set of single nucleon
solutions in the mean meson fields (instead of in vacuum) and for
meson fields around their suitably chosen  non-zero mean values
(instead of around the vacuum). This is a generalization of the
usual quantization procedure around the vacuum, and is equivalent to
it. Instead the loop expansion, we take the expansion scheme in
which terms are classified according to the number of mesons in the
intermediate states. This is something like the Tamm-Dancoff
method\cite{t,d} used in particle physics. Numerical calculations
here will be limited to the approximation in which only one meson
appears in the intermediate state. This is in the spirit of the one
boson exchange (OBE) idea in the traditional nuclear physics. This
procedure relates with the RMFT and with the usual nuclear theory
more close. Here we would check if and in which case the quantum
correction on RMFT may be regarded as a small perturbation.

Section \ref{form} is a formulation of the quantization procedure
for the scalar meson field around its mean value and that for the
nucleon field in mean scalar meson field. Its equivalence with the
usual quantization procedure in vacuum is shown. In section
\ref{correction} we apply this procedure to the $\sigma$-$\omega$
model\cite{w74} for nuclear matter, both the RMFT contribution and
its quantum corrections are derived. Numerical results are given.
Section \ref{sec5} is devoted to  conclusions and discussions.
\section{Formulation}\label{form}
Consider a system consisting of a neutral scalar meson field $\Phi$
and a nucleon field $\Psi$, its Lagrangian density in the nature
units $\hbar =c=1$ is
\begin{eqnarray}{\cal L}&=&\bar{\Psi}\left({\rm
i}\gamma_\mu\partial^\mu
-m\right)\Psi+\frac12\partial_\mu\Phi\partial^\mu\Phi-\frac12m_\sigma^2\Phi^2\nonumber\\
&&+g_\sigma\bar\Psi\Phi\Psi\, ,\label{1}\end{eqnarray} Let us
introduce an arbitrary classical mean value $\phi$ for the scalar
meson field , which is assumed to be independent of space-time.
Defining
\begin{equation}\Phi'=\Phi-\phi\, ,\label{2}\end{equation}
we write the Lagrangian density (\ref{1}) in the form
\begin{eqnarray}{\cal L}&=&\bar{\Psi}\left({\rm
i}\gamma_\mu\partial^\mu
-m'\right)\Psi+\frac12\partial_\mu\Phi'\partial^\mu\Phi'-\frac12m_\sigma^2\Phi'^2\nonumber\\
&&-\frac12m_\sigma^2\phi^2-g_\sigma C
\Phi'+g_\sigma\bar\Psi\Phi'\Psi\, ,\label{5}
\end{eqnarray}with\begin{eqnarray}m'&=&m-g_\sigma\phi\, ,\\
C&=&\frac{m_\sigma^2}{g_\sigma}\phi
\end{eqnarray}

The quantization of the nucleon field $\Psi$ in (\ref{1}) and
(\ref{5}) seems to be the same. But the sets of eigenfunctions on
which one expands the field operator $\Psi$ are different. In the
former case we expand $\Psi$ in terms of the complete set of
eigenfunctions
$\frac{1}{\sqrt{(2\pi)^3}}u_{s\tau}(\textbf{k})\exp\left({\rm
i}\textbf{k}\cdot\textbf{x}\right)$ and
$\frac{1}{\sqrt{(2\pi)^3}}v_{s\tau}(\textbf{k})\exp\left(-{\rm
i}\textbf{k}\cdot \textbf{x}\right)$ of the single nucleon energy
operator $\vec{\alpha }\cdot (-\mbox{i}\nabla)+\beta m$ in vacuum:
\begin{eqnarray}\Psi(\textbf{x})&=&\int\frac{{\rm
d}^3k}{\sqrt{(2\pi)^3}}\sum_{s\tau}\left[c_{s\tau}(\textbf{k})u_{s\tau}(\textbf{k})
{\rm e}^{{\rm
i}\textbf{k}\cdot\textbf{x}}\right.\nonumber\\&&\left.+d^\dag_{s\tau}(\textbf{k})
v_{s\tau}(\textbf{k}){\rm e}^{-{\rm
i}\textbf{k}\cdot\textbf{x}}\right]\, ,\label{15}\\
\bar\Psi(\textbf{x})&=&\int\frac{{\rm
d}^3k}{\sqrt{(2\pi)^3}}\sum_{s\tau}\left[d_{s\tau}(\textbf{k})\bar{v}_{\textbf{k}s\tau}
{\rm e}^{{\rm i}\textbf{k}\cdot\textbf{x}}\right.\nonumber\\
&&\left.+{c^\dag_{s\tau}}(\textbf{k})\bar {u}_{\textbf{k}s\tau}{\rm
e}^{-{\rm i}\textbf{k}\cdot\textbf{x}}\right]\,
.\label{16}\end{eqnarray} $s$ and $\tau$ are spin and isospin
indices respectively, the nucleon spinor states
$u_{s\tau}(\textbf{k})$ and $v_{s\tau}(\textbf{k})$ satisfy
equations
\begin{eqnarray}&&\left(\vec{\alpha}\cdot \textbf{k}+\beta
m\right)u_{s\tau}(\textbf{k})=\omega(k) u_{s\tau}(\textbf{k})\,
,\\&&\left(\vec{\alpha}\cdot \textbf{k}-\beta
m\right)v_{s\tau}(\textbf{k})=\omega(k) v_{s\tau}(\textbf{k})\,
,\end{eqnarray} with
\begin{eqnarray}&&\omega(k)=\sqrt{k^2+m^2}\, ,\\&&u^\dag_{s\tau}(\textbf{k})
u_{s'\tau'}(\textbf{k})=\delta_{ss'}\delta_{\tau\tau'}\\
&&v^\dag_{s\tau}(\textbf{k}) v_{s'\tau'}(\textbf{k})=
\delta_{ss'}\delta_{\tau\tau'}\, ,\\&&u^\dag_{s\tau}(\textbf{k})
v_{s'\tau'}(\textbf{k})=v^\dag_{s\tau}(\textbf{k})
u_{s'\tau'}(\textbf{k})=0\, ,\end{eqnarray} quantization conditions
are
\begin{eqnarray} &&
c_{s\tau}(\textbf{k})c_{s'\tau'}(\textbf{k}')+c_{s'\tau'}(\textbf{k}')c_{s\tau}(\textbf{k})\nonumber\\
&=&d_{s\tau}(\textbf{k})d_{s'\tau'}(\textbf{k}')+d_{s'\tau'}(\textbf{k}')d_{s\tau}(\textbf{k})\nonumber\\
&=&c_{s\tau}(\textbf{k})d_{s'\tau'}(\textbf{k}')+d_{s'\tau'}(\textbf{k}')c_{s\tau}(\textbf{k})\nonumber
\\&=&c_{s\tau}(\textbf{k})d^\dag_{s'\tau'}(\textbf{k}')+d^\dag_{s'\tau'}(\textbf{k}')c_{s\tau}(\textbf{k})=0\,
,\\&&c_{s\tau}(\textbf{k})c^\dag_{s'\tau'}(\textbf{k}')+c^\dag_{s'\tau'}(\textbf{k}')c_{s\tau}(\textbf{k})\nonumber\\
&=&d_{s\tau}(\textbf{k})d^\dag_{s'\tau'}(\textbf{k}')+d^\dag_{s'\tau'}(\textbf{k}')d_{s\tau}(\textbf{k})\nonumber\\
&=&\delta_{ss'}\delta_{\tau\tau'}\delta(\textbf{k}-\textbf{k}')\,
.\end{eqnarray} The vacuum state $|0\rangle$ is defined to be the
eigenstate of annihilation operators with zero eigenvalues. It means
\begin{eqnarray}
c_{s\tau}(\textbf{k})|0\rangle=d_{s\tau}(\textbf{k})|0\rangle=0\,
.\label{C}\end{eqnarray}  To make the expectation value be zero in
vacuum, the Hamiltonian density is expressed in terms of normal
products, in which annihilation operators $c_{s\tau}(\textbf{k})$
and $d_{s\tau}(\textbf{k})$ always stand on the right of creation
operators $c^\dag_{s'\tau'}(\textbf{k}')$ and
$d^\dag_{s'\tau'}(\textbf{k}')$. The nucleon sector of the
Hamiltonian density is therefore
\begin{eqnarray}{\cal H}_1&=&:\Psi^\dag\left(-{\rm i}\vec{\alpha}\cdot\nabla+\beta
m\right)\Psi
-g_\sigma\Phi\bar\Psi\Psi:\nonumber\\&=&\Psi^\dag\left(-{\rm
i}\vec{\alpha}\cdot\nabla+\beta m\right)\Psi
-g_\sigma\Phi\bar\Psi\Psi\nonumber\\&&+\frac{\gamma}{2\pi^2}\int_0^\infty\left(\sqrt{k^2+m^2}
-g_\sigma\Phi\frac{m}{\sqrt{k^2+m^2}}\right)k^2{\rm d}k
,\nonumber\\\end{eqnarray} products sandwiched between two colons
are defined to be normal. For nuclear matter, $\gamma=4$ (neutrons
and protons with spin up and down), and for neutron matter,
$\gamma=2$.

In the later case,  it is in the classical scalar meson field
$\phi$, we may quantize $\Psi$ in the same way, but have to
substitute
$m',\omega',c'_{s\tau}(\textbf{k}),d'_{s\tau}(\textbf{k})$ and
$|0\rangle'$ for
$m,\omega,c_{s\tau}(\textbf{k}),d_{s\tau}(\textbf{k})$ and
$|0\rangle$ respectively. The nucleon sector of the Hamiltonian
density is therefore
\begin{eqnarray}{\cal H}'_1&=&:\Psi^\dag\left(-{\rm i}\vec{\alpha}\cdot\nabla+\beta
m'\right)\Psi
-g_\sigma\Phi'\bar\Psi\Psi:'\nonumber\\&=&\Psi^\dag\left(-{\rm
i}\vec{\alpha}\cdot\nabla+\beta m\right)\Psi
-g_\sigma\Phi\bar\Psi\Psi\nonumber\\&&+\frac\gamma{2\pi^2}\int_0^\infty\left(\sqrt{k^2+m'^2}
-g_\sigma\Phi'\frac{m'}{\sqrt{k^2+m'^2}}\right)k^2{\rm d}k
.\nonumber\\\end{eqnarray}Products sandwiched between $:$ and $:'$
are defined to be normal in the sense that annihilation operators
$c'_{s\tau}(\textbf{k})$ and $d'_{s\tau}(\textbf{k})$ are on the
right of creation operators $c'^\dag_{s'\tau'}(\textbf{k}')$ and
$d'^\dag_{s'\tau'}(\textbf{k}')$.

The difference between these two expressions gives
\begin{eqnarray}{\cal H}_1&=&{\cal H}'_1+\frac{\gamma}{2\pi^2}\left[\int_0^\infty\left(\sqrt{k^2+m^2}
-\sqrt{k^2+m'^2}\right)k^2{\rm
d}k\right.\nonumber\\&&-g_\sigma\Phi'\int_0^\infty\left(\frac{m}{\sqrt{k^2+m^2}}
-\frac{m'}{\sqrt{k^2+m'^2}}\right)k^2{\rm
d}k\nonumber\\&&\left.-g_\sigma\phi\int_0^\infty\frac{m}{\sqrt{k^2+m^2}}k^2{\rm
d}k\right]\, .
\end{eqnarray}
The integrals have been worked out analytically. Deleting a fourth
degree polynomial in $\phi$ and $\Phi'$ by the renormalization of
scalar meson field energy density, we obtain
\begin{eqnarray} {\cal H}_1&=&{\cal H}'_1-A+g_\sigma B\Phi'\, ,\\ A&=&\frac\gamma{16\pi^2}\left[m'^4\ln\frac{m'}{m}+m^3(m-m')
-\frac{7}{2}m^2(m-m')^2\right.\nonumber\\&&+\left.\frac{13}{3}m(m-m')^3
-\frac{25}{12}(m-m')^4\right],
\\B&=&\frac\gamma{4\pi^2}\left[m'^3\ln\frac{m'}{m}+m^2(m-m')
-\frac{5}{2}m(m-m')^2\right.\nonumber\\&&+\left.\frac{11}{6}(m-m')^3\right].
\end{eqnarray}

The usual quantization of the scalar meson field $\Phi$ and its
canonically conjugated variable $\Pi$ in vacuum is performed by the
expansions
\begin{eqnarray}\Phi(\textbf{x})&=&\int\frac{{\rm
d}^3k}{\sqrt{(2\pi)^32\omega_\sigma(k)}} \left[a(\textbf{k}){\rm
e}^{{\rm
i}\textbf{k}\cdot\textbf{x}}\right.\nonumber\\
&&\left.+ a^\dag(\textbf{k}){\rm e}^{-{\rm
i}\textbf{k}\cdot\textbf{x}}\right],\label{18}\\
\Pi(\textbf{x})&=&\int\frac{{\rm
d}^3k}{\sqrt{(2\pi)^3}}\sqrt{\frac{\omega_\sigma(k)}{2}} {\rm
i}\left[ a^\dag(\textbf{k}){\rm e}^{-{\rm
i}\textbf{k}\cdot\textbf{x}}\right.\nonumber\\
&&\left.-a(\textbf{k}){\rm e}^{{\rm
i}\textbf{k}\cdot\textbf{x}}\right] ,\label{181}\end{eqnarray}
with
\begin{equation}\omega_\sigma(k)=\sqrt{k^2+m^2_\sigma}\;.\end{equation}
Quantization conditions are\begin{eqnarray}
&&a(\textbf{k})a(\textbf{k}')-a(\textbf{k}')a(\textbf{k})=0\, ,
\label{A}\\&&a(\textbf{k})a^\dag(\textbf{k}')-a^\dag(\textbf{k}')a(\textbf{k})=\delta(\textbf{k}-\textbf{k}')\,
.\label{B}\end{eqnarray} Instead, one may quantize the scalar meson
field $\Phi'$ around a classical value $\phi$ by the same procedure.
It is to substitute $\Phi',\Pi'$ and $a'(\textbf{k})$ for $\Phi,\Pi$
and $a(\textbf{k})$ in equations (\ref{18}), (\ref{181}),(\ref{A}),
and (\ref{B}). The relation (\ref{2}) gives
\begin{eqnarray}a(\textbf{k})&=&a'(\textbf{k})+\phi\sqrt{\frac{\omega_\sigma(2\pi)^3}2}
\delta^3(\textbf{k})\label{28}\\
a^\dag(\textbf{k})&=&{a'}^\dag(\textbf{k})+\phi\sqrt{\frac{\omega_\sigma(2\pi)^3}2}
\delta^3(\textbf{k})
\end{eqnarray}
Since the difference of $\Phi$ and $\Phi'$ is an additive c-number,
we have
\begin{eqnarray}&&:\frac12m_\sigma^2\Phi^2:=:\frac12m_\sigma^2\Phi'^2+\frac12m_\sigma^2\phi^2+g_\sigma C\Phi':'\, ,\\
&&:\Pi^2+(\nabla\Phi )^2:=:\Pi'^2+(\nabla\Phi' )^2:'\,
,\end{eqnarray} in which the normal products sandwiched in $:\ :$
are defined in terms of annihilation operators $a(\textbf{k})$ and
creation operators $a^\dag(\textbf{k})$, while the normal products
sandwiched in $:\ :'$ are defined in terms of annihilation operators
$a'(\textbf{k})$ and creation operators $a'^\dag(\textbf{k})$. The
Hamiltonian of the nucleon-scalar meson system is therefore
\begin{eqnarray}H&=&\int\left\{ :\Psi^\dag\left(-{\rm i}\vec{\alpha}\cdot\nabla+\beta
m\right)\Psi
-g_\sigma\bar\Psi\Phi\Psi:\right.\nonumber\\
&&\left.+\frac12:\left[\Pi^2+(\nabla\Phi )^2+m_\sigma^2\Phi^2\right]
:\right\}{\rm d}^3x\nonumber\\
&=&\int \left\{:\Psi^\dag\left(-{\rm i}\vec{\alpha}\cdot\nabla+\beta
m'\right)\Psi:'-A
\right.\nonumber\\&&+\frac{1}{2}:\left[\Pi'^2+(\nabla\Phi'
)^2+m_\sigma^2\Phi'^2\right]:'+\frac12m_\sigma^2\phi^2\nonumber\\&&\left.+g_\sigma
\left[\left(B+C\right)\Phi'-:\bar\Psi\Phi'\Psi:'\right]\right\}{\rm
d}^3x\, ,\label{32}\end{eqnarray} This equation shows the
equivalence of the quantization procedure in and around the mean
meson field with the usual one in and around the vacuum. Write
\begin{eqnarray}H=H^{(0)}+H'\, ,\label{33}\end{eqnarray} with
\begin{eqnarray}H^{(0)}&=&\int\!
\omega'(k)\sum_{s\tau}\left[c_{s\tau}'^\dag(\textbf{k})c_{s\tau}'(\textbf{k})
\!+\!d_{s\tau}'^\dag(\textbf{k})d_{s\tau}'(\textbf{k})\right]{\rm
d}^3k\nonumber\\
&& +\int\omega_\sigma(k)a'^\dag(\textbf{k})a'(\textbf{k}){\rm
d}^3k\,\nonumber\\&&+\int \left(\frac12m_\sigma^2\phi^2-A\right){\rm
d}^3x\\\omega'(k)&=&\sqrt{k^2+m'^2}
,\\H'&=&g_\sigma\int\left[(B+C)\Phi'-:\bar{\Psi}\Phi'\Psi:'\right]{\rm
d}^3x\, .\label{36}\end{eqnarray}The unperturbed Hamiltonian
$H^{(0)}$ depends only on the mean value $\phi$ of the scalar meson
field, no single scalar meson appear in it. It is therefore the RMFT
Hamiltonian for the nucleon-scalar meson system. Since $A$ is the
change of the nucleon vacuum energy due to the appearance of $\phi$,
the term containing it represents a quantum effect. The usual RMFT
discards this term, but Chin's RHA takes it. RHA is therefore an
extended RMFT, or shortly the ERMFT. The perturbation $H'$ contains
the quantum correction due to the fluctuation of the scalar meson
field around its mean value $\phi$ and the nucleon-scalar meson
interaction. The theory considering both $H^{(0)}$ and $H'$ is
therefore a quantized RMFT, or shortly the QRMFT.
\section{Quantum corrections of the Walecka $\sigma$-$\omega$ model for nuclear
matter}\label{correction} The zeroth order ground state of an
iso-symmetric static uniform nuclear matter is defined by the
eigen-state $|\eta \rangle $ of $H^{(0)}$ with nucleons filled in
the positive energy Fermi sea from the bottom up to the Fermi
surface of momentum $\eta$, and without any single scalar meson. The
contribution of $H'$ is considered by perturbation up to the second
order.

A vector meson field is needed to stabilize the nuclear matter. The
Walecka $\sigma$-$\omega$ model treats the nuclear matter as a
nucleon-scalar meson-vector meson field system by RMFT. $\sigma$ is
an iso-singlet scalar meson with a mass about several hundred MeV to
be determined by the comparison of the theoretical calculation with
nuclear data, while $\omega$ is an iso-singlet vector meson with a
mass of 783MeV. The contribution of the vector meson on the energy
density of a static uniform nuclear matter is $g_\omega
n^2/2m_\omega$, in which $g_\omega$ is the $\omega$-nucleon coupling
constant, $m_\omega$ is the $\omega$ meson mass, and
$n=\gamma\eta^3/6\pi^2$ is the nucleon number density of nuclear
matter.This contribution comes from a second order perturbation of
the nucleon-vector meson coupling.  The energy per-nucleon of the
nuclear matter in units of $c=\hbar=m=1$ is then
\begin{eqnarray}
\epsilon
=\epsilon_1+\epsilon_2+\frac{g^2_\omega}{2m^2_\omega}n-\frac An
-\frac{g^2_\sigma\left(\tilde{n}-B-C\right)^2}{2m^2_\sigma n}\;
 ,\label{anes}
\end{eqnarray}
in which
\begin{eqnarray}
\epsilon_1\!=\!\frac{3}{4}\!\left[\!\!\left(\!1+\frac{\chi^2}{2\eta^2}\!\right)\!\!
\!\sqrt{\eta^2\!+\chi^2}\!-\!\!\frac{\chi^4}{2\eta^3}
\!\ln\!\!\left(\!\frac{\eta}{|\chi |}+\sqrt{1+\frac{\eta^2}{\chi^2}}
\right)\!\!\right] \label{nes}\end{eqnarray} is the average energy
of a nucleon in the nuclear matter, with $\chi =m'/m=1-g\phi /m$ to
be the effective nucleon mass in unit of its free mass
$m$;\begin{eqnarray}\epsilon_2=\frac{3\pi^2m^2_\sigma}{\gamma\eta^3g^2_\sigma}(1-\chi)^2
=\frac{m^2_\sigma}{2ng^2_\sigma}(1-\chi)^2 \label{un}\end{eqnarray}
is the energy of the mean scalar meson field per nucleon;
\begin{equation}\tilde{n}=\frac{\gamma\chi}{4\pi^2}\left[\eta\sqrt{\eta^2+\chi^2}-\chi^2
\ln\left(\frac\eta{\left|\chi\right|}+\sqrt{1+\frac{\eta^2}{\chi^2}}\right)\right]\end{equation}
is the scalar nucleon number density of the nuclear matter; with
\begin{eqnarray}A&=&\frac\gamma{16\pi^2}\left[\chi^4\ln\chi+\left(1-\chi\right)-\frac72\left(1-\chi\right)^2
\right.\nonumber\\
&&\left.+\frac{13}3\left(1-\chi\right)^3-\frac{25}{12}\left(1-\chi\right)^4\right]
,\\B&=&\frac
\gamma{4\pi^2}\left[\chi^3\ln\chi+\left(1-\chi\right)-\frac52\left(1-\chi\right)^2
\right.\nonumber\\
&&\left.+\frac{11}6\left(1-\chi\right)^3\right] ,\\
C&=&\frac{m^2_\sigma(1-\chi)}{g^2_\sigma}.
\end{eqnarray}

The sum of the first three terms on the right of (\ref{anes}) is
exactly the usual RMFT result considered by Walecka\cite{w74}, while
the last two terms are quantum corrections. The fourth term comes
from the change of vacuum energy due to the appearance of the mean
field $\phi$, and therefore shows the renormalized vacuum
fluctuation. The sum of the first four terms has been considered by
Chin in his relativistic Hartree approximation (RHA)\cite{c1}. As we
explained before, it is an extended version of the RMFT. The last
term is our new, it comes from the second order perturbation of
$H'$, and is somewhat complex. There is a square of a sum with three
terms. Expanding the square, one obtains nine terms. The one with
$\tilde{n}^2$ comes from the interaction between nucleons in
positive energy Fermi sea by the exchange of a scalar meson in
medium, it is the quantum of the field $\Phi'$. This is an effect of
the in-medium OBEP, and is shown in FIG.\ref{fig1}.
\begin{figure}
\begin{fmffile}{phi}
\begin{fmfgraph*}(30,15)
\fmfleft{i} \fmfright{o} \fmf{fermion,right}{i,v1,i}
\fmf{dashes}{v1,v2} \fmf{fermion,left}{v2,o,v2}
\end{fmfgraph*}\end{fmffile} \caption{Feynman
diagram}\label{fig1}\end{figure}
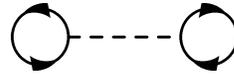 The term with $B^2$ comes from the
change of interaction energy between negative energy nucleons in
vacuum by exchanging a scalar meson in medium, due to the appearance
of the mean field $\phi$. The term with $C^2$ comes from the change
of the scalar meson field vacuum. Other terms show the mixing and
interference of these three effects. Altogether, they show typical
quantum effects. The sum of these five terms is the result of QRMFT
in the second order perturbation approximation.
\section{NUMERICAL RESULTS}\label{sec4}
The energy $\epsilon$ per-nucleon  is a function of $\chi$ and
$\eta$.  $\chi$ is determined by the condition
\begin{equation}
\frac{\partial\epsilon}{\partial\chi}=0\label{44}
\end{equation}
of the energy minimization. It makes $\epsilon$  be a function of
$\eta$ alone, and therefore be a function $\epsilon(n)$ of nucleon
number density $n$. This is the nuclear equation of state. The model
parameters are reduced to two independent dimensionless parameters
$\alpha_\sigma\equiv m^2g^2_\sigma/m^2_\sigma$ and
$\alpha_\omega\equiv m^2g^2_\omega/m^2_\omega$. They are chosen to
reproduce the binding energy $b=15.986$MeV per-nucleon and the
equilibrium density $3/(4\pi r_0^3)$ of nuclear matter  with
$r_0=1.175$fm\cite{ms}.

The resulting parameters are listed in the TABLE \ref{tab1}.
\begin{table}\caption{Parameters and calculated properties of
iso-symmetric nuclear matter at equilibrium
density}\begin{tabular}{cccccc} \hline & $\displaystyle
\alpha_\sigma$& $\displaystyle \alpha_\omega$ &&$\displaystyle\chi$&
$K$(MeV)\\\hline RMFT&362.6&278.0&&0.538&554\\
RHA&230.3&149.0&&0.730&456\\
QRMFT&230.3&149.0&&0.730&456\\\hline\end{tabular}\label{tab1}\end{table}
The first line is given by Walecka's RMFT, which is obtained by
approximating $\epsilon$ by the sum of its first three terms in
(\ref{anes}) before to be substituted into the variational condition
(\ref{44}). The second line is given by Chin's RHA, which is
obtained by approximating $\epsilon$ by the sum of its first four
terms in (\ref{anes}) before to be substituted into the variational
condition (\ref{44}). The third line is given by our QRMFT, which is
obtained by substituting the whole expression (\ref{anes}) for
$\epsilon$ into the variational condition (\ref{44}). The sets of
parameters are adjusted to reproduce the same set of nuclear data
listed above. We see that quantum corrections notably change the
parameters. The last two columns show two of the most important data
calculated by corresponding methods. The low density equation of
state is almost determined by the compression modulus
$K$\cite{bs83}. One of the shortcomings of the Walecka's RMFT is
that it gives a too large $K$ value. Various methods, such as the
introduction of non-linear terms\cite{bs83}, have been introduced to
overcome this shortcoming. However we see that it is noticeably
reduced by the quantum corrections, without introducing any
additional parameter. Another important property is the effective
mass $\chi$, which determines the energy dependence of the optical
potential. The value 0.730 in last two lines is consistent with the
reasonable range $0.65\le\chi\le0.75$\cite{fp81}, while that of the
Walecka's RMFT is too small.

\begin{figure}\includegraphics[width=9cm]{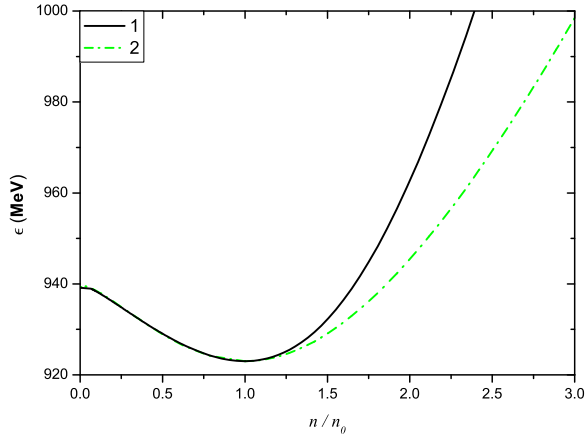}
\caption{Equation of state for iso-symmetric uniform nuclear matter,
given by RMFT (line 1) and by RHA or QRMFT (line 2)}\label{fig2}
\end{figure}

Looking at TABLE \ref{tab1}, one amazedly sees that the results of
RHA and QRMFT are the same. This is also shown in FIG.
\ref{fig2}.
The equations of state for RHA and QRMFT are the same. It means that
the quantum correction on RHA given by the last term of (\ref{anes})
is practically zero, and therefore the RHA solution of the problem
defined by the Lagrangian (\ref{1}) is an excellent zeroth order
approximation for the next order quantum correction.

\begin{figure}\includegraphics[width=9cm]{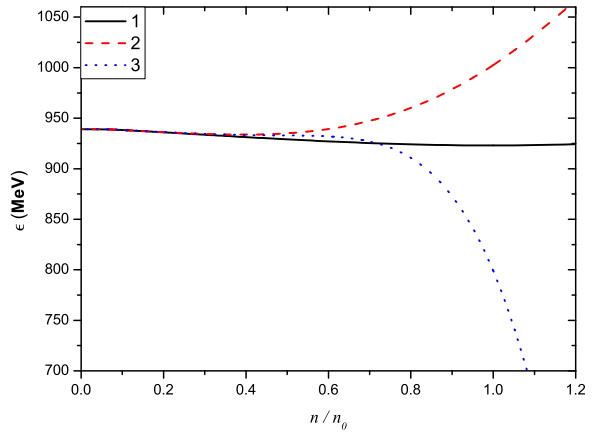}
\caption{Line 1: The RMFT equation of state. Line 2: The equation of
state obtained by adding the fourth term of (\ref{anes}) on line 1.
Line 3: The equation of state obtained by adding the fifth term of
(\ref{anes}) on line 2.}\label{fig3}
\end{figure}

On the other hand, Fig.
\ref{fig3}
 shows that the quantum
corrections on RMFT given by the last two terms of (\ref{anes}) are
quite big, and make the equation of state qualitatively distorted.
The series of quantum corrections seems not converge. It means that
the Walecka's RMFT solution of the problem defined by the Lagrangian
(\ref{1}) is not a good zeroth order approximation for the next
order quantum corrections.

Of course, these results are for the problem defined by the
Lagrangian (\ref{1}). The situation should be checked case by case.
The smallness of the correction on RHA is due to the cancelation
between $\tilde{n},B$ and $C$. In the problem considered above, the
cancelation is almost complete. However, the completeness will not
be always. There is evidence showing that if the Lagrangian
(\ref{1}) is generalized in a way with the term
$\frac12m_\sigma^2\Phi^2$ being substituted by
\begin{equation}U(\Phi)=\frac12m_\sigma^2\Phi^2\left(1+a_1\Phi+a_2\Phi^2\right), \end{equation}
the quantum corrections on RHA may be non-zero but still small, for
non-zero $a_1$ and $a_2$.
\section{Conclusions and discussions}\label{sec5}
Shortly speaking, our conclusions are: \begin{enumerate}\item The
RHA covers the leading part of quantum field theory for nuclear
systems, but the simple RMFT does not. \item The residual
interaction $H'$ in Eq. (\ref{36}) may be considered by perturbation
theory, with the expansion scheme used in traditional nuclear
physics.
\item QRMFT, which we proposed here, is therefore an easy and
practical way for handling nuclear problems until quark degrees of
freedom become important.\end{enumerate}

In our formalism, the mean field $\phi$ is a free parameter. The
equivalence (\ref{32}) is true for its any value. This makes us have
right to choose its value so that the solution of our problem
becomes easy and accurate. RMFT, RHA and QRMFT chose it in different
ways, and then obtained different accuracies and convergence
properties for their solutions.  Among them the QRMFT is the best.
In this way we solve the problem of non-convergence, which the loop
expansion procedure also encountered before\cite{fps}. Different
value of $\phi$ means different vacuum in the nuclear matter. The
theory with a change of vacuum must be non-perturbative. The vacuum
$|0\rangle'$ in nuclear matter satisfies the condition
\begin{equation}a'(k)|0\rangle'=0.\end{equation}
By (\ref{28}) we see
\begin{equation}a(k)|0\rangle'=\phi\sqrt{\frac{\omega_\sigma(2\pi)^3}2}
\delta^3(\textbf{k})|0\rangle'.\end{equation} It means $|0\rangle'$
is a coherent state of the original scalar mesons. Scalar mesons
develop a Bose condensation in nuclear matter.

Beside the scalar meson, any kinds of mesons may develop Bose
condensation in nuclear matter under suitable condition. The most
often talked is the $\pi$-condensation. We have even proven that the
$\pi$-condensation may develop in the Walecka model\cite{z4,wyw}. In
non-uniform nuclear systems, for example the finite nuclei, the
vector meson condensation may develop. In these cases, the quantum
fluctuation of the condensed mesons should also be considered. The
meson condensation is described by a classical meson field, and the
quantum fluctuation is described by its quantum corrections. The
method developed here may be useful for considering these
corrections.

\end{document}